\documentclass[a4paper]{jpconf}

\usepackage{mathtools}
\usepackage{amssymb}
\usepackage[hidelinks]{hyperref}
\usepackage[capitalise]{cleveref}
\usepackage[square,sort&compress,numbers]{natbib}
\usepackage{graphicx}
\usepackage{xspace}
\usepackage[nolist]{acronym}
\usepackage{titlesec}

\graphicspath{{./img/}}

\newcommand{\python}[0]{\texttt{Python}\xspace}
\newcommand{\cpp}[0]{\texttt{C++}\xspace}
\newcommand{\njet}[0]{\texttt{NJet}\xspace}
\newcommand{\sherpa}[0]{\texttt{Sherpa}\xspace}
\newcommand{\eigen}[0]{\texttt{Eigen}\xspace}

\newcommand{\incite}[1]{Ref.~\cite{#1}}

\titlespacing*{\section}{0pt}{10pt}{0pt}

\begin{document}

\begin{acronym}
    \acro{IR}{infrared}
    \acro{ME}{matrix element}
    \acro{ML}{machine learning}
    \acro{MC}{Monte Carlo}
    \acro{NN}{neural network}
    \acro{SM}{Standard Model}
    \acro{LO}{leading order}
    \acro{NLO}{next-to-leading order}
    \acro{NNLO}{next-to-next-to-leading order}
    \acro{QCD}{quantum chromodynamics}
    \acro{PDF}{parton distribution function}
\end{acronym}

\title{Optimising hadronic collider simulations using amplitude neural networks}
\author{Ryan Moodie}
\address{
    Institute for Particle Physics Phenomenology,
    Ogden Centre for Fundamental Physics,
    Department of Physics,
    University of Durham,
    South Road,
    Durham,
    DH1 3LE,
    UK
}
\ead{ryan.i.moodie@durham.ac.uk}

\begin{abstract}
    Precision phenomenological studies of high-multiplicity scattering processes at collider experiments present a substantial theoretical challenge and are vitally important ingredients in experimental measurements.
    Machine learning technology has the potential to dramatically optimise simulations for complicated final states.
    We investigate the use of neural networks to approximate matrix elements, studying the case of loop-induced diphoton-plus-jets production through gluon fusion.
    We train neural network models on one-loop amplitudes from the \njet \cpp library and interface them with the \sherpa Monte Carlo event generator to provide the matrix element within a realistic hadronic collider simulation.
    Computing some standard observables with the models and comparing to conventional techniques, we find excellent agreement in the distributions and a reduced total simulation time by a factor of thirty.
\end{abstract}

\vspace{-8pt}
\section{Introduction}

With the increasing Large Hadron Collider dataset driving ever more precise experimental measurements, \ac{SM} predictions for high-multiplicity scattering at hadronic colliders form a vital part of precision phenomenology studies.
Currently, these calculations mainly rely on automated numerical codes \cite{degrande:2018neu} to calculate \acp{ME} up to high multiplicities, including tree-level real corrections at \ac{NLO} and double-real corrections at \ac{NNLO}, and one-loop real-virtual corrections at \ac{NNLO}.
These codes have been a theoretical revolution, particularly at one-loop, but the evaluation time is relatively high.
Due to the high dimensionality of the phase space, these real-type corrections are often the computational bottleneck in higher-order calculations.

Following recent advances in precision \ac{QCD}, there has been a flurry of activity around \ac{NNLO} \ac{QCD} corrections to diphoton-plus-jet production, including full-colour two-loop amplitudes \cite{agarwal:2021vdh} and leading-colour \ac{NNLO} distributions \cite{chawdhry:2021hkp}.
In the loop-induced gluon fusion channel, the full-colour two-loop amplitudes were computed \cite{badger:2021imn}, leading to full-colour \ac{NLO} distributions \cite{badger:2021ohm}; it was found that the opening of this new channel enhances the mixed quark- and gluon-initiated \ac{NNLO} prediction \cite{chawdhry:2021hkp} in certain regions, disrupting the usual perturbative convergence.
Also for diphoton production through gluon fusion, the three-loop amplitudes were calculated \cite{bargiela:2021wuy}, making available the final remaining piece for its \ac{NNLO} corrections.
Therefore, we study the loop-induced class of processes with two photons and many gluons as they are extremely relevant for current phenomenology.

\Ac{ML} technology has found a wealth of application in high energy physics \cite{feickert:2021ajf}.
We employ the ensemble \ac{NN} model of \incite{badger:2020uow}, which studied $e^+e^-$ annihilation to jets, to emulate the gluon-initiated diphoton-plus-jets \acp{ME} within a full \ac{MC} event generator simulation.
This tests the methodology against the additional complexity of hadronic collider simulations, including \ac{PDF} convolution and variable centre-of-mass scales, complex phase space cuts and jet clustering, and phase space sampling optimisation methods of integrators.

This contribution is organised as follows.
We first discuss the gluon-initiated diphoton-plus-jets amplitudes and their implementations within the \cpp \njet library \cite{badger:2012pg,badger:2013vpa} which were used for training.
We then describe the phase space partitioning used to handle \ac{IR} divergent regions.
Next, we present the architecture of the \acp{NN} used.
Then, we discuss the simulation pipeline and interface of the \ac{NN} model to the \sherpa \ac{MC} event generator \cite{bothmann:2019yzt}.
Finally, we study the performance of the model compared to the original amplitude library for $gg\to\gamma\gamma gg$ and present some distributions before concluding.

This contribution is based on \incite{aylett-bullock:2021hmo}.
Our code is publicly available \cite{n3jet_diphoton}.

\section{Amplitudes}

\begin{figure}[t]
    \begin{minipage}{0.24\textwidth}
        \begin{center}
            \vspace{5.25em}
            \includegraphics[width=\textwidth]{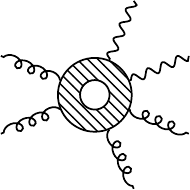}
            \caption{
                \label{fig:ggyygg_1l}
                Diagram of $gg\to\gamma\gamma gg$ ($N=6$) at \acs{LO}.
                The photons couple to an internal quark loop.
            }
        \end{center}
    \end{minipage}
    \hfill
    \begin{minipage}{0.74\textwidth}
        \begin{center}
            \includegraphics[width=0.66\textwidth]{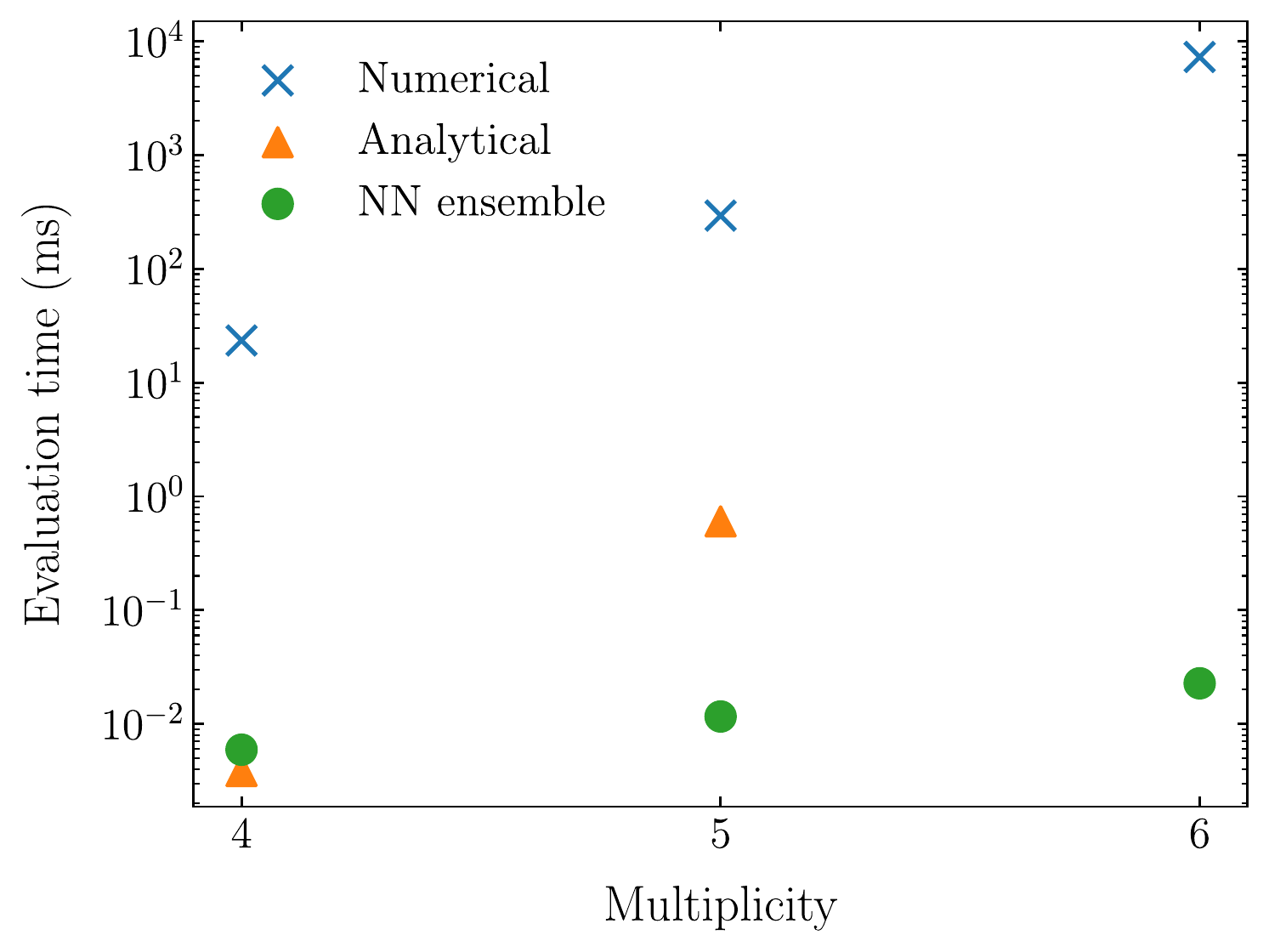}
            \caption{
                \label{fig:timing-single}
                Single-threaded CPU evaluation time of the \ac{ME} for a single phase space point.
                The value is the mean of 100 evaluations over a random phase space.
                Results are shown for available implementations at various multiplicities, including numerical and analytical evaluations using \njet and inference on the \ac{NN} model.
            }
        \end{center}
    \end{minipage}
\end{figure}

As there is no vertex coupling gluons to photons in the \ac{SM}, diphoton-plus-jets production through gluon fusion (\cref{fig:ggyygg_1l}) is loop induced.
The \ac{LO} process is $\mathcal{O}({\alpha_s}^{N-2})$ for multiplicity $N$, appearing at \ac{NNLO} in the perturbative expansion of the combined quark- and gluon-initiated process.
We study the channels with only gluons and photons in the external particles, $gg\to\gamma\gamma+n\times g$.
These proceed through a quark loop at \ac{LO}.

Conventional event generator simulations optimise virtual corrections in \ac{NLO} calculations by learning the phase space of the \ac{LO} process and using this to sample the virtual contribution.
This technique fails for loop-induced processes, where the expensive one-loop amplitude has no tree-level process to optimise the phase space on.
Therefore, new methods are required to improve the efficiency of integrating these channels at high multiplicity.

We use the one-loop-squared \acp{ME} from the \njet library as the targets for our \ac{NN} emulation.
These include two classes of amplitudes: an automated numerical setup for arbitrary multiplicity; and hard-coded analytical expressions for $N\in\{4,5\}$.
The numerical implementation obtains the diphoton-plus-jets amplitudes by summing permutations of pure-gluon primitive amplitudes \cite{deflorian:1999tp}, which are themselves based on generalised unitary \cite{badger:2008cm} and integrand reduction \cite{ossola:2006us}.
While completely automated, evaluation time and numerical stability are increasingly difficult to control.
The hard-coded implementations offer compact analytical expressions with extremely fast and stable evaluation, although they are unavailable for higher multiplicity.
The $N=5$ result is obtained through a finite field reconstruction \cite{peraro:2019svx}.
The evaluation timings of these methods are compared to the \ac{NN} model in \cref{fig:timing-single}.

\section{Phase space partitioning}

Training a single \ac{NN} over the entire phase space results in a poor fit, especially at higher multiplicity \cite{badger:2020uow}.
This is caused by regions where the \ac{ME} becomes \ac{IR} divergent, which arise from soft ($s_i$) and collinear ($c_{ij}$) emissions.
These singularities are regulated with cuts, but the \ac{ME} is rapidly varying in local regions, which causes problems for the global fit.
Therefore, we train a separate \ac{NN} on each of the \ac{IR} structures of the phase space.

We first partition the phase space into a non-divergent region, $\mathcal{R}_{\textrm{non-div}}$, and a divergent region, $\mathcal{R}_{\textrm{div}}$.
Points that pass a cut, $\text{min}(\{s_{ij}/s_{12} : i,j\in\{1,\ldots,N\}\})<y$, are included in $\mathcal{R}_{\textrm{div}}$.
The threshold $y$ must be tuned to discriminate points of similar scales into each region, while having sufficient points in $\mathcal{R}_{\textrm{div}}$ to train on.

We then sub-divide $\mathcal{R}_{\textrm{div}}$ according to the decomposition of the FKS subtraction scheme \cite{frederix:2009yq}.
We define a set of FKS pairs, $\mathcal{P}_{\text{FKS}} = \left\{ (i,j) : s_i \lor s_j \lor c_{ij} \right\}$, corresponding to the $\binom{N}{2} - 1$ singular configurations, which includes redundancy (App.~B of \incite{badger:2020uow}).
Each pair is assigned a partition function,
$
\mathcal{S}_{ij} = 1 / \left( s_{ij} \sum_{j,k\in \mathcal{P}_{\text{FKS}}} 1/s_{jk} \right),
$
which smoothly isolates that divergence on multiplication with the \ac{ME}.

We train a \ac{NN} on $\left|\mathcal{A(\boldsymbol{p})}\right|^2$ for $\boldsymbol{p}\in\mathcal{R}_{\textrm{non-div}}$, and a \ac{NN} on each of the partition-function-weighted \acp{ME},
$
\left\{\mathcal{S}_{ij} \left|\mathcal{A(\boldsymbol{p})}\right|^2 \,:\, i,j\in\mathcal{P}_{\text{FKS}} \,;\, \boldsymbol{p}\in\mathcal{R}_{\textrm{div}}\right\}.
$
We reconstruct the complete \ac{ME} in $\mathcal{R}_{\textrm{div}}$ by summing the weighted \acp{ME},
$
\left|\mathcal{A}\right|^2 = \sum_{i,j\in \mathcal{P}_{\text{FKS}}} \mathcal{S}_{ij} \left|\mathcal{A}\right|^2.
$
This ensemble of \acp{NN}, referred to as the model, can be used to accurately infer the \ac{ME} over the complete phase space, $\mathcal{R}_{\textrm{non-div}}\cup\mathcal{R}_{\textrm{div}}$.

Note that increasing the cut $y$, which increases the proportion of points in $\mathcal{R}_{\textrm{div}}$, incurs a performance penalty due to the higher cost of inferring over several \acp{NN} in $\mathcal{R}_{\textrm{div}}$ compared to the single \ac{NN} in $\mathcal{R}_{\textrm{non-div}}$.

\section{Model architecture}

Although using fine-tuned architectures for each configuration (process, cuts, etc.) would provide optimal performance, this would be prohibitively expensive.
We use a general setup as the gains of specialised \ac{NN} optimisation are beyond the scope of this pioneering work, performing hyperparameter optimisation on the $gg\to\gamma\gamma g$ process.

Each \ac{NN} uses a fully-connected architecture, a standard choice for a regression problem, parameterised using the Keras \python interface \cite{keras} to the TensorFlow \ac{ML} library \cite{tensorflow}.
There are $4\times N$ input nodes: one for each component of each momentum in the phase space point.
The three hidden layers are comprised of 20, 40, and 20 nodes respectively, all with hyperbolic-tangent activation functions.
There is a single output node with a linear activation function, which returns the approximation of the \ac{ME}.
We find that this is a sufficient number of layers and nodes to learn the \acp{ME}, while remaining economical for computational performance.

We train with a mean-squared-error loss function, using Adam-optimised stochastic gradient descent \cite{adam}.
We expect that the model will learn the mean of the target distribution using this loss function (App.~A of \incite{badger:2020uow}).
The number of training epochs is determined by Early Stopping regularisation, with a patience of 100 epochs to mitigate the effects of the limited size of $\mathcal{R}_\text{div}$ that may appear in the validation set.
We use 32-bit floating-point numbers throughout.

\section{Pipeline}

Our \ac{ML} pipeline for the $gg\to\gamma\gamma gg$ results presented is: generate the training and validation datasets by running \sherpa with \njet while uniformly sampling phase space; train the model; infer on the model to estimate the \acp{ME} during event generation with \sherpa, using the original integration grid.

Input data consists of a list of phase space points, $p_i^{\mu}\in\mathbb{R}^{4N}$, and the corresponding colour- and helicity-summed one-loop-squared \ac{ME}, $\left|\mathcal{A}\right|^2\in\mathbb{R}$.
Phase space sampling is determined by the integrator, meaning the training is optimal only for a specific integrator.
The data is extracted from a run of the integrator, generating 100k points which are split 4:1 into training and validation datasets.
A 3M point testing dataset is produced by a second run of the integrator with a different random number seed and used to evaluate model performance.

We perform inference on an ensemble of twenty models, each of which have different random weight initialisation and shuffled training and validation datasets.
We take as the result the mean of the ensemble, with the standard error providing the precision/optimality error \cite{badger:2020uow}.

While training was performed using \python, event generators are generally written in \cpp.
To use the model within a simulation, we wrote a \cpp inference code and a bespoke \cpp interface for \sherpa.
The weights of the trained models are written to file and read by the inference code at runtime; the library \eigen \cite{eigenweb} is used to perform efficient linear algebra on the CPU.
The interface can also be used to call \cpp amplitude libraries directly; we use this to interface \njet to \sherpa to generate the datasets, which is performed with 64-bit floats.

\Acp{PDF} are provided by LHAPDF \cite{buckley:2014ana} using the NNPDF3.1 set \texttt{NNPDF31\_nlo\_as\_0118} \cite{ball:2017nwa}.
Cuts are adapted from those in \incite{badger:2013ava}.
Analysis is performed using \texttt{Rivet} \cite{bierlich:2019rhm} with an adapted reference analysis script \cite{aaboud:2017vol}.

\section{Results}

\begin{figure}[t]
    \begin{minipage}{0.49\textwidth}
        \begin{center}
            \includegraphics[width=\textwidth]{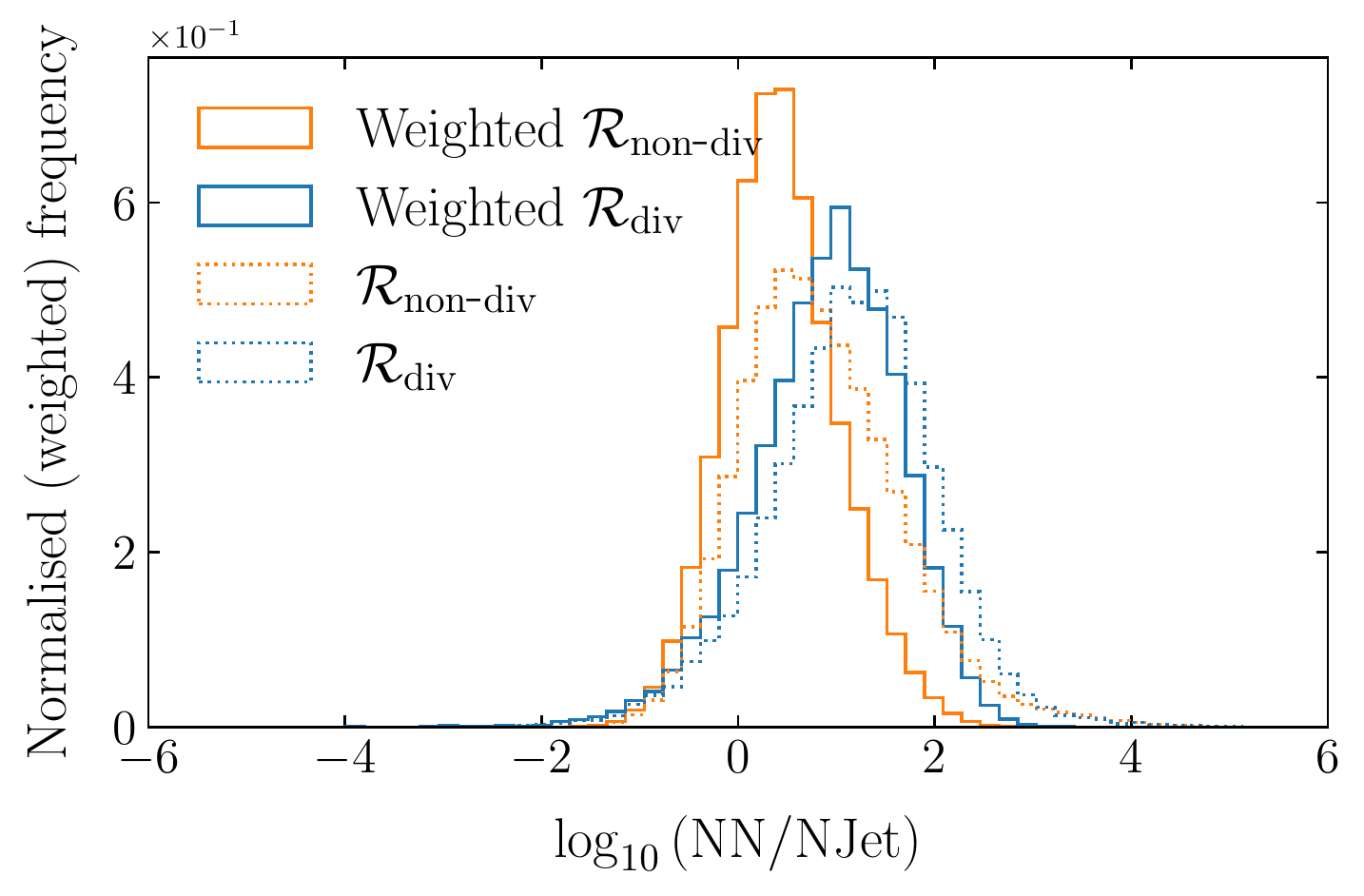}
            \caption{
                \label{fig:error-plot}
                \acs{PDF}-weighted and unweighted histograms by region of the logarithm of the ratio between the \ac{ME} returned by the model and \njet for each point in a 1M subset of the training data for $gg\to\gamma\gamma gg$. The region cut is $y=10^{-3}$ and $\mathcal{R}_{\textrm{div}}$ contains 2.4\% of the points.
            }
        \end{center}
    \end{minipage}
    \hfill
    \begin{minipage}{0.49\textwidth}
        \begin{center}
            \vspace{0.2em}
            \includegraphics[width=\textwidth]{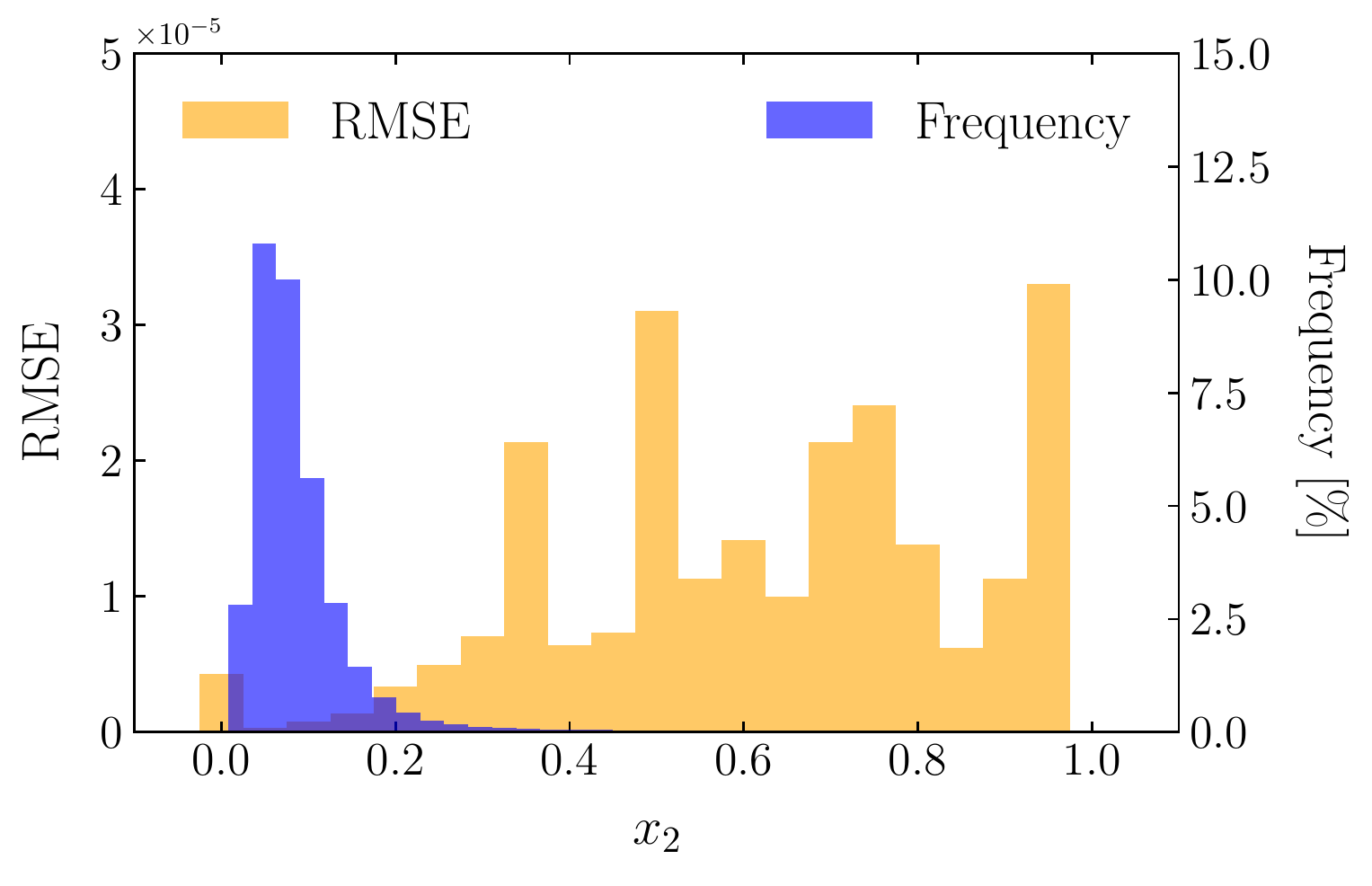}
            \caption{
                \label{fig:x2-error-plot}
                Histogram of the root mean squared error (RMSE) of the model compared to \njet for a univariate slice of phase space in $x_2$, the momentum fraction of the second incoming gluon, for $gg\to\gamma\gamma g$.
                Also shown are the points in the training dataset, binned in $x_2$.
            }
        \end{center}
    \end{minipage}
\end{figure}

Comparing the output of the trained $gg\to \gamma\gamma gg$ model to the amplitude library value by unweighted point-by-point ratio in \cref{fig:error-plot}, we see a peaked and approximately symmetric error distribution with a shifted mean in both regions.
Both region histograms have a similar mean, indicating comparable accuracy, with $\mathcal{R}_{\textrm{non-div}}$ performing slightly better.
The distributions are fairly broad.
$\mathcal{R}_{\textrm{non-div}}$ shows a slight tail on the right, which arises from points near the cutoff $y$.

Despite the per-point agreement being somewhat poor, the total cross section is found to be in agreement, with $\sigma_\text{\ac{NN}} = (45 \pm 6) \times 10^{-7} \thinspace \mathrm{pb}$ (precision/optimality error) and $\sigma_\mathrm{NJet} = (49 \pm 5) \times 10^{-7} \thinspace \mathrm{pb}$ (\ac{MC} error).
\Cref{fig:x2-error-plot} shows that the regions that are sampled the most due to the shape of the gluon \ac{PDF} are those that have the lowest error.
In addition, the \ac{PDF}-weighted histograms in \cref{fig:error-plot} are more narrowly peaked and closer to being unit-centred than the unweighted histograms.
Thus, the agreement in the total cross section is much better than for point-by-point comparison because poorly performing points fall in \ac{PDF}-suppressed regions.
This indicates that the accuracy of distributions inferred with the model is dependent on the choice of process, cuts, and observable.

\begin{figure}[t]
    \begin{minipage}{0.49\textwidth}
        \begin{center}
            \includegraphics[width=\textwidth]{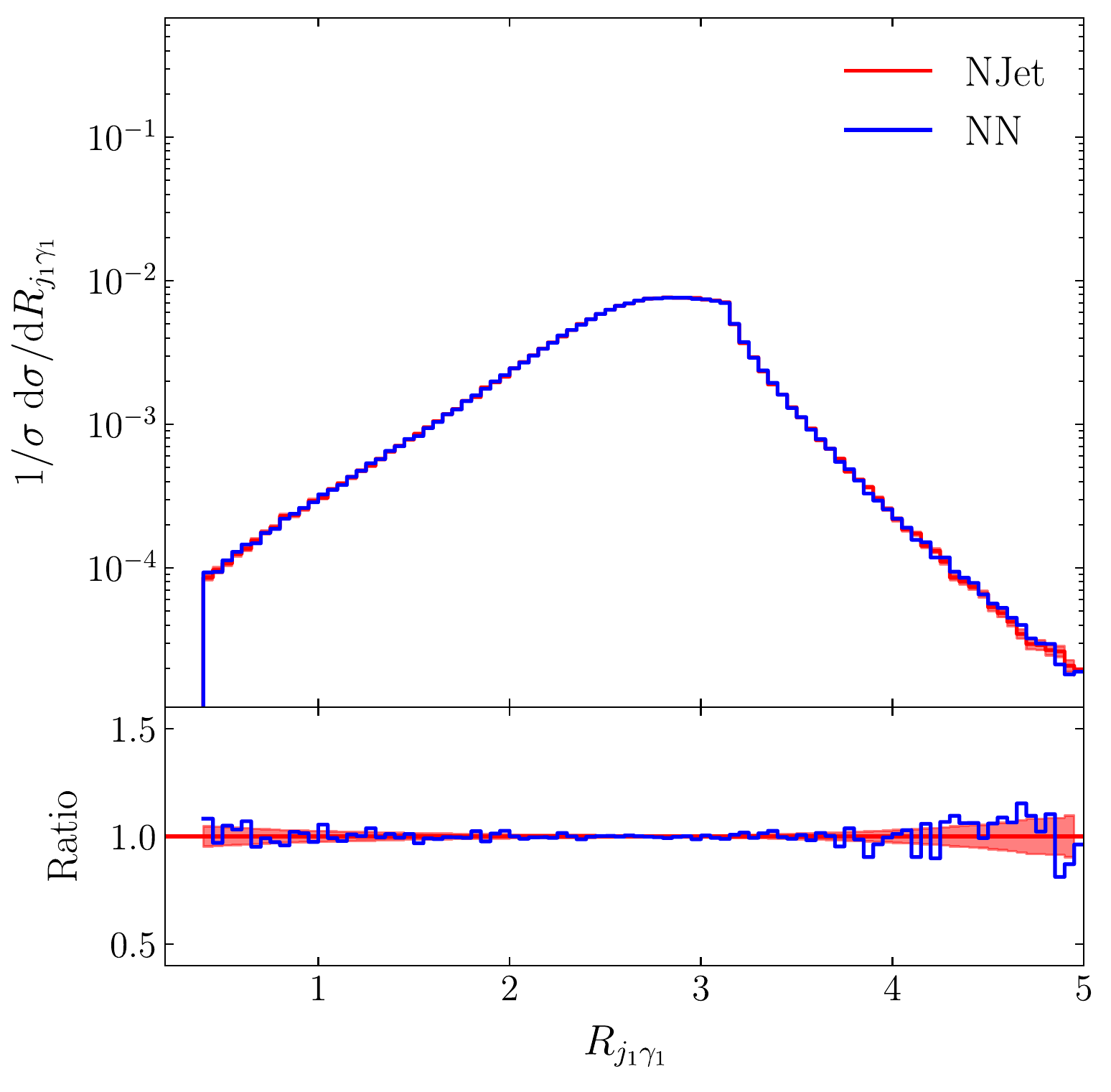}
        \end{center}
    \end{minipage}
    \hfill
    \begin{minipage}{0.49\textwidth}
        \begin{center}
            \includegraphics[width=\textwidth]{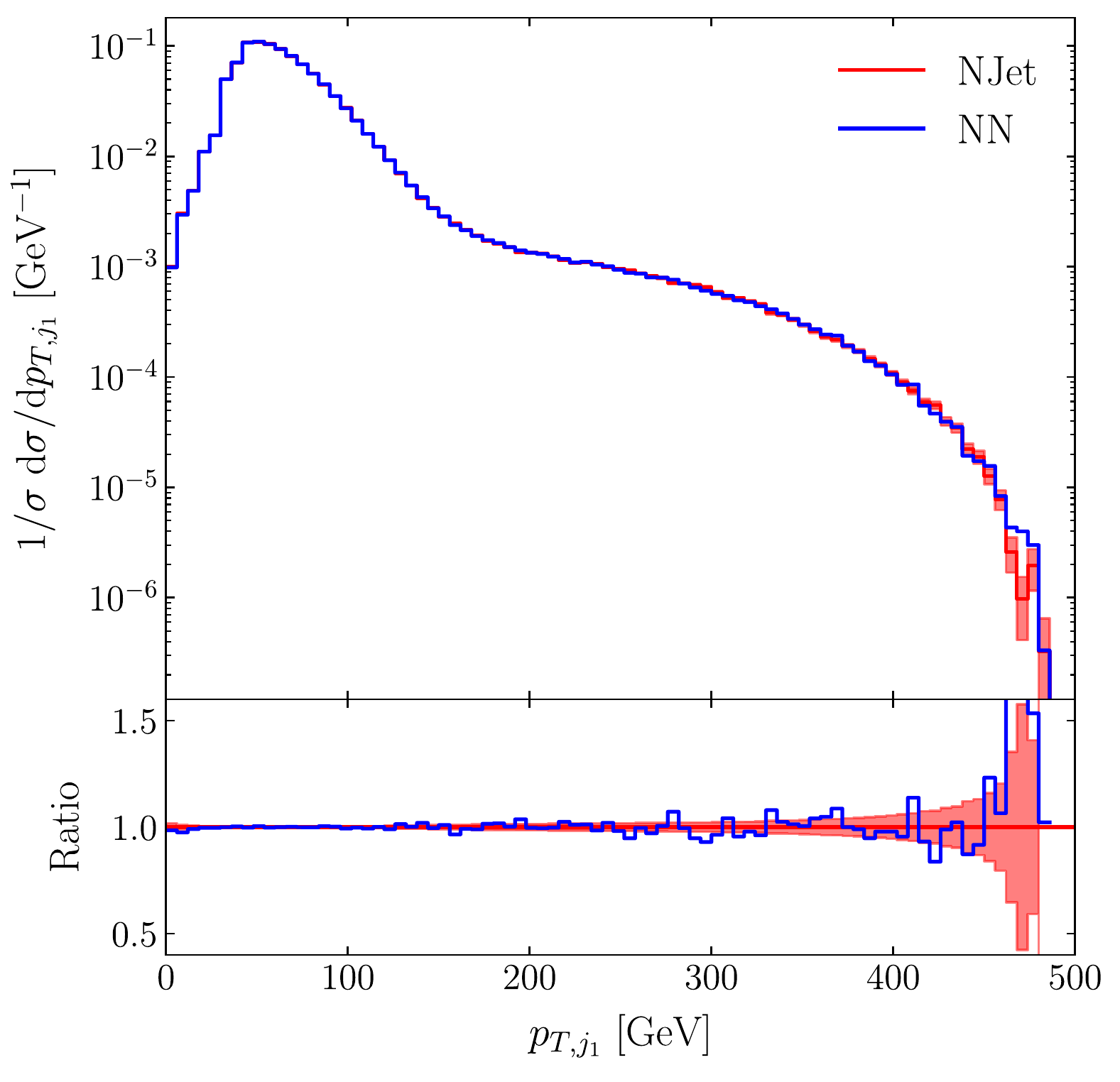}
        \end{center}
    \end{minipage}
    \caption{
        \label{fig:xs}
        Differential normalised cross sections for $gg\to\gamma\gamma gg$, comparing \njet (\ac{MC} error) to the model (precision/optimality error), in R-separation between the hardest jet and photon (left) and the transverse momentum of the hardest jet (right).
        Refer to \incite{aylett-bullock:2021hmo} for definitions of observables and cuts, and further distributions.
    }
\end{figure}

\cref{fig:xs} shows excellent agreement between the distributions obtained from the model and \njet for two differential slices of phase space.
There are some fluctuations in the tails although they appear statistical rather than systematic and the model predictions mostly remain within the \njet \ac{MC} error bands.
Normalised \ac{NN} uncertainties are negligible compared to the \ac{MC} error.

In \incite{aylett-bullock:2021hmo}, we also demonstrates how agreement can be improved in $\mathcal{R}_{\textrm{div}}$ by reweighting event weights by the ratio of the emulated and true \acp{ME} at known points from the training data, as well as showing good performance for $gg\to\gamma\gamma g$ when relaxing cuts at inference compared to training.

Subsequent to this work, the authors of \incite{maitre:2021uaa} achieve improved per-point agreement at tree-level by exploiting the factorisation properties of \acp{ME}.

\section{Conclusion}

We extend previous work which pioneered the emulation of \acp{ME} with \acp{NN}, studying these techniques for the first time within a full hadronic collider simulation.
We focus on loop-induced diphoton-plus-jets production via gluon fusion.
The difficulties introduced by \ac{IR} behaviour are tamed by partitioning the phase space as prescribed by FKS subtraction.
We provide a general interface for trained models to \sherpa.

We find that amplitude \ac{NN} models provide an efficient and general framework for optimising high-multiplicity observables at hadronic colliders.
Agreement in differential distributions is excellent.
As the cost of inference is negligible compared to the amplitude library call in training, the speed up in total simulation time (including training) compared to conventional methods is given by the ratio of the number of points used for inference and training, $N_{\mathrm{infer}}/N_{\mathrm{train}}$.
For this study, this gave a factor of thirty, although for studies with higher statistics or coverage of multiple cut configurations, the factor would be much greater.

\ack{
    I would like to thank Joseph Aylett-Bullock and Henry Truong for useful discussions, and Simon Badger for comments on the draft manuscript. I am supported by UKRI-STFC ST/S505365/1 and ST/P001246/1.
}

\bibliography{proc}

\providecommand{\newblock}{}
\begin{thebibliography}{10}
\expandafter\ifx\csname url\endcsname\relax
  \def\url#1{{\tt #1}}\fi
\expandafter\ifx\csname urlprefix\endcsname\relax\def\urlprefix{URL }\fi
\providecommand{\eprint}[2][]{\url{#2}}

\bibitem{degrande:2018neu}
Degrande C, Hirschi V and Mattelaer O 2018 {\em Ann. Rev. Nucl. Part. Sci.\/}
  {\bf 68} 291--312

\bibitem{agarwal:2021vdh}
Agarwal B, Buccioni F, von Manteuffel A and Tancredi L 2021 {\em Phys. Rev.
  Lett.\/} {\bf 127} 262001 (\textit{Preprint}
  \href{http://arxiv.org/abs/2105.04585}{2105.04585})

\bibitem{chawdhry:2021hkp}
Chawdhry H~A, Czakon M, Mitov A and Poncelet R 2021 {\em JHEP\/} {\bf 09} 093
  (\textit{Preprint} \href{http://arxiv.org/abs/2105.06940}{2105.06940})

\bibitem{badger:2021imn}
Badger S {\em et~al.\/} 2021 {\em JHEP\/} {\bf 2021} (\textit{Preprint}
  \href{http://arxiv.org/abs/2106.08664}{2106.08664})

\bibitem{badger:2021ohm}
Badger S, Gehrmann T, Marcoli M and Moodie R 2022 {\em Phys. Lett. B\/} {\bf
  824} 136802 (\textit{Preprint}
  \href{http://arxiv.org/abs/2109.12003}{2109.12003})

\bibitem{bargiela:2021wuy}
Bargiela P, Caola F, von Manteuffel A and Tancredi L 2022 {\em JHEP\/} {\bf 02}
  153 (\textit{Preprint} \href{http://arxiv.org/abs/2111.13595}{2111.13595})

\bibitem{feickert:2021ajf}
Feickert M and Nachman B 2021  (\textit{Preprint}
  \href{http://arxiv.org/abs/2102.02770}{2102.02770})

\bibitem{badger:2020uow}
Badger S and Bullock J 2020 {\em JHEP\/} {\bf 06} 114 (\textit{Preprint}
  \href{http://arxiv.org/abs/2002.07516}{2002.07516})

\bibitem{badger:2012pg}
Badger S, Biedermann B, Uwer P and Yundin V 2013 {\em Comput. Phys. Commun.\/}
  {\bf 184} 1981--1998 (\textit{Preprint}
  \href{http://arxiv.org/abs/1209.0100}{1209.0100})

\bibitem{badger:2013vpa}
Badger S, Biedermann B, Uwer P and Yundin V 2014 {\em J. Phys. Conf. Ser.\/}
  {\bf 523} 012057 (\textit{Preprint}
  \href{http://arxiv.org/abs/1312.7140}{1312.7140})

\bibitem{bothmann:2019yzt}
Bothmann E {\em et~al.\/} (Sherpa) 2019 {\em SciPost Phys.\/} {\bf 7} 034
  (\textit{Preprint} \href{http://arxiv.org/abs/1905.09127}{1905.09127})

\bibitem{aylett-bullock:2021hmo}
Aylett-Bullock J, Badger S and Moodie R 2021 {\em JHEP\/} {\bf 2021}
  (\textit{Preprint} \href{http://arxiv.org/abs/2106.09474}{2106.09474})

\bibitem{n3jet_diphoton}
Aylett-Bullock J and Moodie R 2021 n3jet\_diphoton v1
  \url{https://gitlab.com/JosephPB/n3jet_diphoton}

\bibitem{deflorian:1999tp}
de~Florian D and Kunszt Z 1999 {\em Phys. Lett. B\/} {\bf 460} 184--188
  (\textit{Preprint}
  \href{http://arxiv.org/abs/hep-ph/9905283}{hep-ph/9905283})

\bibitem{badger:2008cm}
Badger S~D 2009 {\em JHEP\/} {\bf 01} 049 (\textit{Preprint}
  \href{http://arxiv.org/abs/0806.4600}{0806.4600})

\bibitem{ossola:2006us}
Ossola G, Papadopoulos C~G and Pittau R 2007 {\em Nucl. Phys. B\/} {\bf 763}
  147--169 (\textit{Preprint}
  \href{http://arxiv.org/abs/hep-ph/0609007}{hep-ph/0609007})

\bibitem{peraro:2019svx}
Peraro T 2019 {\em JHEP\/} {\bf 07} 031 (\textit{Preprint}
  \href{http://arxiv.org/abs/1905.08019}{1905.08019})

\bibitem{frederix:2009yq}
Frederix R, Frixione S, Maltoni F and Stelzer T 2009 {\em JHEP\/} {\bf 10} 003
  (\textit{Preprint} \href{http://arxiv.org/abs/0908.4272}{0908.4272})

\bibitem{keras}
Chollet F {\em et~al.\/} 2015 Keras \url{https://github.com/fchollet/keras}

\bibitem{tensorflow}
Abadi M {\em et~al.\/} 2015 {TensorFlow} \url{https://www.tensorflow.org/}

\bibitem{adam}
Kingma D~P and Ba J 2015 {\em 3rd International Conference for Learning
  Representations\/} (\textit{Preprint}
  \href{http://arxiv.org/abs/1412.6980}{1412.6980})

\bibitem{eigenweb}
Guennebaud G, Jacob B {\em et~al.\/} 2010 Eigen v3
  \url{https://eigen.tuxfamily.org}

\bibitem{buckley:2014ana}
Buckley A {\em et~al.\/} 2015 {\em Eur. Phys. J. C\/} {\bf 75} 132
  (\textit{Preprint} \href{http://arxiv.org/abs/1412.7420}{1412.7420})

\bibitem{ball:2017nwa}
Ball R~D {\em et~al.\/} (NNPDF) 2017 {\em Eur. Phys. J. C\/} {\bf 77} 663
  (\textit{Preprint} \href{http://arxiv.org/abs/1706.00428}{1706.00428})

\bibitem{badger:2013ava}
Badger S, Guffanti A and Yundin V 2014 {\em JHEP\/} {\bf 03} 122
  (\textit{Preprint} \href{http://arxiv.org/abs/1312.5927}{1312.5927})

\bibitem{bierlich:2019rhm}
Bierlich C {\em et~al.\/} 2020 {\em SciPost Phys.\/} {\bf 8} 026
  (\textit{Preprint} \href{http://arxiv.org/abs/1912.05451}{1912.05451})

\bibitem{aaboud:2017vol}
Aaboud M {\em et~al.\/} (ATLAS) 2017 {\em Phys. Rev. D\/} {\bf 95} 112005
  (\textit{Preprint} \href{http://arxiv.org/abs/1704.03839}{1704.03839})

\bibitem{maitre:2021uaa}
Ma\^{i}tre D and Truong H 2021 {\em JHEP\/} {\bf 11} 066 (\textit{Preprint}
  \href{http://arxiv.org/abs/2107.06625}{2107.06625})

\end{thebibliography}

\end{document}